\documentclass[12pt,twoside]{article}
\usepackage{amsmath}
\usepackage{amsfonts}

\newcommand{\bra}[1]{\langle #1 |}
\newcommand{\ket}[1]{|#1\rangle}

\newcommand{\tr}{\textrm{tr}}

\newcommand{\V}{\mathcal{V}}
\newcommand{\Op}{\mathcal{O}}
\newcommand{\R}{\tilde{\rho}}
\begin{document}
\title{Approach to Thermal Equilibrium in the Caldeira-Leggett Model}
\author{Fethi M Ramazanoglu\\
\small Department of Physics,\\
\small Princeton University,\\
\small Princeton, NJ-08544\\[-0.25in]}
\date{}
\maketitle \pagestyle{myheadings}
\markboth{F.M.Ramazanoglu}{Approach to Thermal Equilibrium} \thispagestyle{empty}
\begin{abstract}
\noindent We provide an explicit analytical calculation that shows the asymptotic approach of the one dimensional Caldeira-Leggett model to thermal equilibrium in the high temperature and weak coupling limit. We investigate a free particle and a harmonic oscillator system, using both the Lindblad and the non-Lindblad type master equations for each case, and show that thermal equilibrium is reached exactly for the free particle and aproximately for the harmonic oscillator, irrespective of the initial preparation of the system. We also generalize our calculation to higher dimensions.
\end{abstract}

\section{Introduction}
The Caldeira-Leggett model~\cite{ref:CLoriginal} is a simple system-reservoir model that can explain the basic aspects of dissipation in solid state physics, and in the high temperature and weak coupling limit, can also account for quantum Brownian motion~\cite{ref:Weiss}\cite{Vacchini2000}\cite{Vacchini2005}. It consists of a particle, which is also called ``the system'', that interacts with a heat bath of simple harmonic oscillators through a linear term.

In this study, we first derive the exact solution of the Caldeira-Leggett master equation in a novel way and then use our results to investigate the long time behaviour of the density matrix and the approach of the system to thermal equilibrium.

We first briefly introduce the Caldeira-Leggett master equation in both the Lindblad and the non-Lindblad form, and examine the stationary solutions of the Lindblad form and their relationship to the thermal equilibrium density matrix (Sec.~\ref{sec:setting_master}). After introducing our basic mathematical tools, we explicitly solve the time evolution of the density matrix for a free particle system using the Lindblad form of the master equation, and investigate its behaviour in the long time limit (Sec.\ref{sec:free_evolution}). We then study the differences that would arise from using the non-Lindblad form of the master equation (Sec.~\ref{sec:non-Lindblad}). Next, we repeat all the previous calculations for the harmonic oscillator system (Sec.~\ref{sec:oscillator}) and finally show that all our results are generalizable to higher dimensions (Sec.~\ref{sec:higher_dimensions}). We end with a summary of our results and some comments (Sec.~\ref{sec:conclusions}).

\section{Master Equation and the Stationary States} \label{sec:setting_master}

Caldeira-Leggett model is defined by the following Hamiltonian
\begin{equation}
H = \overbrace{\frac{1}{2m}p^2 + V(q)}^{H_S} +\ \overbrace{q^2 \sum_n \frac{\kappa_n^2}{2m_n\omega_n^2}}^{H_c} +\ \overbrace{ \sum_n \left( \frac{1}{2m_n}p_n^2+\frac{1}{2}m_n \omega_n^2q_n^2\right)}^{H_B} -\ \overbrace{q \sum_n \kappa_n q_n}^{H_I}
\end{equation}
where $q$, $p$ are the position and the momentum operators of the system and $\{q_n\}$, $\{p_n\}$ are the position and momentum operators of the bath oscillators, respectively. $H_S$ is the system hamiltonian, $H_c$ is a counterterm for renormalization, $H_I$ is the interaction term and $H_B$ is the hamiltonian for the reservoir. For our study, the reservoir oscillators are described by an Ohmic spectral density with a Lorentz-Drude cutoff function
\begin{equation}
J(\omega) =\frac{2m\gamma}{\pi} \omega \frac{\Omega^2}{\Omega^2+\omega^2}.
\end{equation}

In the high temperature, weak coupling quantum Brownian motion case in which we are interested, this Hamiltonian leads to the following master equation for the reduced density matrix of the system~\cite{ref:BandP}:
\begin{align}
\frac{d}{dt}\rho_t = -\frac{i}{\hbar}[H_S,\rho_t] -\frac{2\gamma mk_BT}{\hbar^2}[q,[q,\rho_t]]  -i \frac{\gamma}{\hbar} [q,\{p,\rho_t \}]\ . \label{eq:master_density_nonLindblad}
\end{align}
In order for this time evolution to be a quantum dynamical semigroup, one can add a ``minimally invasive'' term, $ -\frac{\gamma}{8mk_B T} [p,[p,\rho_t]]$, which is negligible compared to the other terms in the high temperature limit and which brings the equation into the Lindblad form~\cite{ref:BandP}
\begin{align}
\frac{d}{dt}\rho_t = -\frac{i}{\hbar}[H_S,\rho_t] -\frac{2 \gamma mk_B T}{\hbar^2}[q,[q,\rho_t]] -\frac{\gamma}{8mk_B T} [p,[p,\rho_t]] -i \frac{\gamma}{\hbar} [q,\{p,\rho_t \}] \label{eq:master_density}
\end{align}
with the relaxation rate $\gamma$ and  the Lindblad operator
\begin{equation}
A=\sqrt{\frac{4mk_BT}{\hbar^2}}x+i\sqrt{\frac{1}{4mk_BT}}p\ ,
\end{equation}
which gives the Lindblad evolution
\begin{equation}
\frac{d}{dt}\rho_t = -\frac{i}{\hbar}[H_S+\frac{\gamma}{2}(qp+pq),\rho_t] + \gamma\left(A\rho_tA^{\dagger} -\frac{1}{2}A^{\dagger}A \rho_t -\frac{1}{2}\rho_tA^{\dagger}A \right)\ .
\end{equation}

Certain conditions have to be met for~(\ref{eq:master_density_nonLindblad}) and~(\ref{eq:master_density}) to be valid~\cite{ref:BandP}:
\begin{enumerate}
    \item The typical time scale over which the state of the system changes appreciably, $\tau_R \sim 1/\gamma$, should be much larger than the typical decay time for the correlation functions of the bath oscillators, $\tau_B$, which translates into:
    \begin{equation}
    \hbar \gamma \ll \min\{ \hbar \Omega, 2\pi k_B T\},
    \end{equation}
    The origin of this condition is related to the approximations that make the master equation Markovian and its details can be found in~\cite{ref:BandP}.
    \item The typical system evolution time, $1/\omega_S$, should be large compared to $\tau_B$
    \begin{equation}
    \hbar \omega_S \ll \min\{ \hbar \Omega, 2\pi k_B T\},
    \end{equation}
    This condition is required for the validity of the introduction of the ``minimally invasive term'' as well as for other steps in the derivation of the master equation. As we expect $p \sim m \omega_S q$ for the typical momentum and position values, the ratio of the momentum double commutator to the position double commutator in~(\ref{eq:master_density}) is at the order $\left(\hbar \omega_S/k_BT\right)^2$, thus it vanishes under this condition, making the Lindblad form of the master equation valid.
\end{enumerate}

The first step in understanding the implications of the master equation is studying the stationary solution, i.e. the solution with $d\rho_t/dt=0$. The general expectation is that, in the long time limit, the density matrix is going to reach this solution irrespective of the initial condition. For the non-Lindblad case, (\ref{eq:master_density_nonLindblad}), and for a potential $V(q)$ whose spatial variations are small, the stationary solution in the position representation  is
\begin{equation}
\bra{q_1}\rho \ket{q_2} \approx N\ \exp \left(-\frac{V((q_1+q_2)/2)}{k_BT}-\frac{mk_BT(q_1-q_2)^2}{2\hbar^2}\right)\ ,
\label{eq:stationary_solution_general}
\end{equation}
where $N$ is a normalization constant~\cite{ref:BandP}. Moreover, for the case of the quadratic potential or the free particle, this equation is exact.

For the free particle, ($V(q)=0$),~(\ref{eq:stationary_solution_general}) gives the exact thermal equilibrium state, which can be obtained transforming the familiar expression
\begin{equation}
\bra{p_1} \rho_{th} \ket{p_2} = \bra{p_1} N e^{-\frac{p^2}{2mk_BT}} \ket{p_2}\left\{
\begin{array}{cl}
\sqrt{\frac{1}{2\pi mk_BT}}\ e^{-\frac{p_1^2}{2mk_BT}}&  p_1 = p_2\\
0 & \text{otherwise}
\end{array} \right.
\end{equation}
into the position representation, using the Fourier transform.

For the harmonic oscillator, $V(q)=\frac{1}{2}m\omega^2q^2$, the thermal equilibrium density matrix is given by~\cite{ref:Landau}
\begin{equation}
\bra{q_1}\rho \ket{q_2} = N\ \exp\left[-\frac{m\omega}{2\hbar\ \tanh \left(\hbar \omega/k_BT\right)}\ (q_1^2+q_2^2) + \frac{m\omega}{\hbar\ \sinh \left( \hbar \omega/k_BT \right)}\ q_1q_2\right]
\label{eq:density_matrix_thermal_HO}
\end{equation}
which clearly does not agree with~(\ref{eq:stationary_solution_general}). The source of this diagreement can be traced by expanding the exponent in~(\ref{eq:density_matrix_thermal_HO}) in powers of $\hbar \omega/k_BT$. To the leading order
\begin{align}
\bra{q_1}\rho \ket{q_2} &= N\ \exp\left[- \frac{mk_BT}{2\hbar^2} \left( \frac{\hbar \omega}{k_BT}\right)^2 \frac{(q_1+q_2)^2}{4} - \frac{mk_BT}{2\hbar^2} \left(1+\frac{1}{12}\left(\frac{\hbar \omega}{k_BT}\right)^2\right) (q_1-q_2)^2\right] \nonumber\\
&=N\ \exp\left[-\frac{m \omega^2}{8k_BT} (q_1+q_2)^2 - \frac{mk_BT}{2\hbar^2} (q_1-q_2)^2\right]\ \exp\left[-\frac{m\omega^2}{24k_BT}(q_1-q_2)^2\right] \ ,
\end{align}
which agrees with~(\ref{eq:stationary_solution_general}) on the diagonal ($q_1=q_2$), ignoring the higher order terms, but does not agree with it if we move away from the diagonal where $q_1-q_2$ is comparable to $q_1+q_2$ in magnitude. This should not be surprising, since the term in the exponent that causes the difference, $-\frac{m\omega^2}{24k_BT}(q_1-q_2)^2$, is at the order $\left(\hbar\omega/k_BT\right)^2$ compared to $\frac{mk_BT}{2\hbar^2} (q_1-q_2)^2$. Also, we expect $m\omega^2 q^2 \sim \hbar \omega$, which means $\frac{m\omega^2}{24k_BT}(q_1-q_2)^2 \sim \frac{\hbar\omega}{k_BT}$. We already stated that our master equations is valid in the regime where $\hbar\omega/k_BT \ll 1$, thus it is expected that such terms are not captured by the stationary state of the non-Lindblad master equation. In other words, by agreeing to use~(\ref{eq:master_density_nonLindblad}) or~(\ref{eq:master_density}) to investigate the Caldeira-Leggett model, we give up the prospect to have sensitivity to any higher order terms in $\hbar \omega/k_BT$

To summarize, the stationary solution of the free particle Caldeira-Leggett master equation in the non-Lindblad case is the exact thermal equilibrium density matrix for the free particle. Thus, reaching the stationary state and reaching the thermal equilibrium are equivalent for this case. However, for the harmonic oscillator, the stationary solution is not the thermal equilibrium, but rather, the two agree on the diagonal in the high temperature limit. High temperature limit is essential to the derivation of the master equation itself, thus, we can say that the stationary solution captures the thermal equilibrium as best as possible for the Caldeira-Leggett master equation.

Our previous discussions suggest that the ``minimally invasive term'' should not affect the physics significantly. We are not going to explicitly calculate the stationary solutions with this term, but our subsequent analysis will make it clear that the long time behaviour is identical to the non-Lindblad case.

\section{The Green's Function}
Different techniques have been used to solve the Caldeira-Leggett model, especially for the harmonic oscillator case~\cite{ref:exact_solution1}\cite{ref:exact_solution2}\cite{ref:exact_solution3}. Among these,~\cite{ref:exact_solution3} is the closest to our approach, but it only discusses the non-Lindblad form of the master equation and cannot be readily generalized to the Lindblad master equation (see appendix~\ref{app:general_treatment} for more discussions of this).

For our analysis, we employed the techniques given in the appendices of~\cite{ref:energy_increase} and~\cite{ref:Bassireference}, where they are used for similar but different purposes. The main feature of our technique will be the use of the Wigner function and the Gaussian ansatz. Details of the following calculations can be found in appendix~\ref{app:Wigner_master}.

The characteristic function associated to the Wigner function is defined as
\begin{align}
\tilde{\rho}_t(k,x) &= \tr \left( \rho_t e^{\frac{i}{\hbar}(kq+xp)} \right) \nonumber\\
&= \int du\ e^{\frac{i}{\hbar}xu} \bra{u-\frac{k}{2}} \rho_t \ket{u+\frac{k}{2}} \nonumber\\
&= \int dy\ e^{\frac{i}{\hbar}ky} \bra{y+\frac{x}{2}} \rho_t \ket{y-\frac{x}{2}}\ ,
\label{eq:wigner_definition}
\end{align}
where $\left\{ \ket{u} \right\}$, $\left\{ \ket{y} \right\}$ are the momentum and position bases, respectively. The inversion formulae in the momentum and position basis are
\begin{align} \label{eq:inversion_formulae}
\bra{p_1} \rho_t \ket{p_2} &= \frac{1}{2\pi \hbar} \int dx\ e^{-\frac{i}{\hbar} x(\frac{p_1+p_2}{2})}\ \tilde{\rho}_t (p_2-p_1,x) \nonumber\\
\bra{q_1} \rho_t \ket{q_2} &= \frac{1}{2\pi \hbar} \int dk\ e^{-\frac{i}{\hbar} k(\frac{q_1+q_2}{2})}\ \tilde{\rho}_t (k,q_1-q_2)\ ,
\end{align}
respectively. After some algebraic manipulations, the time evolution for $\tilde{\rho}_t (k,x)$ for the case of $V(q)=0$ is given by
\begin{align}
\frac{\partial}{\partial t} \tilde{\rho}_t (k,x) =\frac{1}{m}\ k \frac{\partial}{\partial x} \R_t (k,x) -\frac{2 \gamma mk_BT}{\hbar^2}\ x^2 \R_t (k,x) -\frac{\gamma}{8mk_B T}\ k^2 \R_t(k,x) -2 \gamma\ x \frac{\partial}{\partial x}\R_t(k,x) \label{eq:master_wigner}
\end{align}

We will employ the Green's function method to solve~(\ref{eq:master_wigner})
\begin{equation}
\R_t(k,x) = \int dk_0 dx_0\ \tilde{G}(k,x,t;k_0,x_0,0) \R_0(k_0,x_0), \label{eq:greens_integral}
\end{equation}
where $\tilde{G}(k,x,t;k_0,x_0,0)$ is the Green's function which is defined as the solution of (\ref{eq:master_wigner}) that satisfies the initial condition
\begin{equation}
\lim_{t \rightarrow 0} \tilde{G}(k,x,t;k_0,x_0,0) = \delta(k-k_0) \delta(x-x_0).
\end{equation}

The key observation is that, under~(\ref{eq:master_wigner}), initially gaussian states remain gaussian. If we make the ansatz
\begin{equation}
\R_t(k,x) = \exp \{ -c_1k^2-c_2kx-c_3x^2-ic_4k-ic_5x -c_6\},
\label{eq:gaussian_ansatz}
\end{equation}
then the master equation leads to
\begin{multline}
-\dot{c_1}k^2-\dot{c_2}kx-\dot{c_3}x^2-i\dot{c_4}k-i\dot{c_5}x -\dot{c_6}\\
= \frac{1}{m} k(-c_2k-2c_3x-ic_5)-\frac{2 \gamma mk_BT}{\hbar^2}x^2 -\frac{\gamma}{8mk_B T}k^2 -2\gamma x (-c_2k-2c_3x-ic_5)\ .
\end{multline}
Upon equating the coefficients of the independent terms, we have the following system of ordinary differential equations:
\begin{eqnarray}
\begin{split}
   \dot{c}_1 (t) &= \frac{c_2 (t)}{m}+ \frac{\gamma}{8mk_B T} \\
   \dot{c}_2 (t) &=  \frac{2c_3 (t)}{m}-2\gamma c_2 (t)\\
   \dot{c}_3 (t) &= \frac{2\gamma mk_BT}{\hbar^2}-4\gamma  c_3 (t)\\
   \dot{c}_4 (t) &=  \frac{c_5 (t)}{m}\\
   \dot{c}_5 (t) &=  -2\gamma c_5 (t)\\
   \dot{c}_6 (t) &=  0\ ,
\end{split}
\label{eq:SHO_c_i}
\end{eqnarray}
which are readily solved using basic techniques:
\begin{equation}
   \begin{split}
      c_1 (t)&= c_1 (0)+c_2 (0)\frac{\Gamma_{t}}{2m\gamma}+c_3(0)\frac{\Gamma_{t}^2}{4m^2\gamma^2}-\frac{k_BT}{8\hbar^2m\gamma^2}(\Gamma_{t}^2 +2\Gamma_{t})+\left( \frac{k_BT}{2\hbar^2m\gamma} +\frac{\gamma}{8mk_BT} \right) t\\
      c_2 (t)&=  c_2 (0)e^{-2\gamma t}+c_3 (0)\frac{\Gamma_{t} e^{-2\gamma t}}{m\gamma}+\frac{ k_BT}{2\hbar^2 \gamma} \Gamma_{t}^2\\
      c_3 (t)&= \frac{mk_BT}{2\hbar^2}+\left(c_3 (0)-\frac{mk_BT}{2\hbar^2} \right)e^{-4\gamma t}\\
      c_4 (t)&=  c_4 (0)+c_5 (0)\frac{\Gamma_{t}}{2m\gamma}\\
      c_5 (t)&= c_5 (0)e^{-2\gamma t} \\
      c_6 (t)&= c_6 (0),
\end{split}
\end{equation}
where $\Gamma_t = 1-e^{-2\gamma t}$.

An initial gaussian of the form
\begin{equation} \label{eq:gaussian_initial}
   \tilde{\rho}_0^{k_0x_0,\epsilon\eta} (k,x)
   =\frac{1}{\pi\sqrt{\epsilon\eta}} e^{-\frac{1}{\epsilon} (k-k_0)^2} e^{-\frac{1}{\eta} (x-x_0)^2}
\end{equation}
has the limit
\begin{equation}
\tilde{\rho}_0^{k_0x_0,\epsilon\eta} (k,x) \xrightarrow[{\epsilon,\eta\rightarrow 0}]{}\delta (k-k_0)\delta (x-x_0)\ ,
\end{equation}
and evolves in time to
\begin{multline}
   \tilde{\rho}_t^{k_0x_0,\epsilon\eta}
   (k,x)=\frac{1}{\pi\sqrt{\epsilon\eta}}\ e^{-\frac{1}{\epsilon}
     (k-k_0)^2}\
e^{-\frac{1}{\eta}\left[x_0- (xe^{-2\gamma t}+\frac{\Gamma_{t} k}{2m\gamma})  \right]^2}\\ \times
   e^{-\left( \frac{k_BT}{2\hbar^2m\gamma} +\frac{\gamma}{8mk_BT} \right) k^2 t }\
e^{\frac{mk_BT}{2\hbar^2}\left[k^2 \frac{\Gamma_t^2+2\Gamma_t}{4m^2\gamma^2}
     - kx\frac{\Gamma_{t}^2}{m\gamma}-x^2 (1-e^{-4\gamma t})\right]}\ .
\end{multline}

We find the Green's function given in~\cite{Vacchini2005} by taking the $\epsilon \to 0, \eta \to 0$  limit:
\begin{multline}
\tilde{G}(k,x,t;k_0,x_0,0) = \delta(k_0-k) \delta\left( x_0- (xe^{-2\gamma t}+\frac{\Gamma_{t} k}{2m\gamma}) \right) \ \ \ \\
\times e^{-\left( \frac{k_BT}{2\hbar^2m\gamma} +\frac{\gamma}{8mk_BT} \right) k^2 t }\ e^{\frac{m k_BT}{2\hbar^2}\left[k^2 \frac{\Gamma_t^2+2\Gamma_t}{4m^2\gamma^2}- kx\frac{\Gamma_{t}^2}{m\gamma}-x^2 (1-e^{-4\gamma t})\right]}\ . \label{eq:free_greens_function}
\end{multline}

\section{Time Evolution of the Free Particle System} \label{sec:free_evolution}

The previous section set up all the background we need to solve for the time dependence of the density matrix. We first insert~(\ref{eq:free_greens_function}) into~(\ref{eq:greens_integral}) to find $\R_t(k,x)$
\begin{align}
\R_t(k,x) = e^{-\left( \frac{k_BT}{2\hbar^2m\gamma} +\frac{\gamma}{8mk_BT} \right) k^2 t }\ e^{\frac{m k_BT}{2\hbar^2}\left[k^2 \frac{\Gamma_t^2+2\Gamma_t}{4m^2\gamma^2}- kx\frac{\Gamma_{t}^2}{m\gamma}-x^2 (1-e^{-4\gamma t})\right]}
\R_0 \left(k,xe^{-2\gamma t}+\frac{\Gamma_{t} k}{2m\gamma} \right)\ . \label{eq:wigner_exact}
\end{align}
%
%
%


In the long time limit, we expect to find that the density matrix will evolve to the stationary solution, which, as we have shown, is diagonal in the momentum representation. So, we will solve the time evolution of the density matrix in this representation:
\begin{align}
\bra{p_1} \rho_t \ket{p_2} &= \frac{1}{2\pi \hbar} \int dx\ e^{-\frac{i}{\hbar} x(\frac{p_1+p_2}{2})} \tilde{\rho}_t (p_2-p_1,x) \nonumber\\
&= \frac{1}{2\pi \hbar} e^{-\left( \frac{k_BT}{2\hbar^2m\gamma} +\frac{\gamma}{8mk_BT} \right) (p_2-p_1)^2 t }\ e^{\frac{(\Gamma_t^2 +2\Gamma_t)k_BT}{8\hbar^2m\gamma^2} (p_2-p_1)^2}\  \nonumber\\
& \ \ \times \int dx\ e^{-\frac{i}{\hbar} x(\frac{p_1+p_2}{2})} e^{\frac{mk_BT}{2\hbar^2}\left[- \frac{\Gamma_{t}^2(p_2-p_1) }{m\gamma}x- (1-e^{-4\gamma t}) x^2\right]}\ \R_0 (p_2-p_1,\ xe^{-2\gamma t} + \frac{\Gamma_t (p_2-p_1)}{2m\gamma} ) \nonumber\\
&= R(t)\ +\nonumber\\
&   \frac{1}{2\pi \hbar} e^{-\left( \frac{k_BT}{2\hbar^2m\gamma} +\frac{\gamma}{8mk_BT} \right) (p_2-p_1)^2 t }\ e^{\frac{(\Gamma_t^2 +2\Gamma_t)k_BT}{8\hbar^2m\gamma^2} (p_2-p_1)^2}\ \R_0(p_2-p_1, \frac{\Gamma_t (p_2-p_1)}{2m\gamma}) \nonumber\\
& \ \  \times \int dx\ e^{-\frac{(1-e^{-4\gamma t})mk_BT}{2\hbar^2} x^2 -\left(\frac{i (p_1+p_2)}{2 \hbar} + \frac{ \Gamma_t^2k_B T}{2\hbar^2 \gamma} (p_2-p_1) \right) x}\nonumber\\
&= R(t)\ + \nonumber\\
& \sqrt{\frac{1}{2\pi mk_BT}} \frac{1}{\sqrt{1-e^{-4\gamma t}}} e^{-\left( \frac{k_BT}{2\hbar^2m\gamma} +\frac{\gamma}{8mk_BT} \right) (p_2-p_1)^2 t }\ e^{\frac{(\Gamma_t^2 +2\Gamma_t)k_BT}{8\hbar^2m\gamma^2} (p_2-p_1)^2}\ \nonumber\\
& \hspace{30mm} \times e^{\frac{\hbar^2}{2(1-e^{-4\gamma t})mk_BT} \left(\frac{i (p_1+p_2)}{2 \hbar} + \frac{ \Gamma_t^2k_B T}{2\hbar^2 \gamma} (p_2-p_1) \right)^2}\  \R_0(p_2-p_1, \frac{\Gamma_t (p_2-p_1)}{2m\gamma}) \label{eq:density_matrix_evolution}
\end{align}
where
\begin{align}
R(t) &=\frac{1}{2\pi \hbar} e^{-\left( \frac{k_BT}{2\hbar^2m\gamma} +\frac{\gamma}{8mk_BT} \right) (p_2-p_1)^2 t }\ e^{\frac{(\Gamma_t^2 +2\Gamma_t)k_BT}{8\hbar^2m\gamma^2} (p_2-p_1)^2}\ \nonumber\\
& \qquad \times \int dx\ \bigg\{ e^{-\frac{(1-e^{-4\gamma t})mk_BT}{2\hbar^2} x^2 -\left(\frac{i (p_1+p_2)}{2 \hbar} + \frac{ \Gamma_t^2k_B T}{2\hbar^2 \gamma} (p_2-p_1) \right) x} \nonumber\\
&\hspace{50mm} \left[ \R_0(p_2-p_1,xe^{-2\gamma t} + \frac{\Gamma_t (p_2-p_1)}{2m\gamma}) - \R_0(p_2-p_1, \frac{\Gamma_t (p_2-p_1)}{2m\gamma}) \right] \bigg\}.
\end{align}
As long as $\R_0(k,x)$ is bounded and well behaved around $(p_2-p_1, \frac{ (p_2-p_1)}{2m\gamma})$, $R(t)$ vanishes in the long time limit.

Then, the main observation is that as $t \to \infty$, $\bra{p_1} \rho_t \ket{p_2}$ vanishes except for the case $p_2-p_1=0$, due to the term $\exp\left[-\left( \frac{k_BT}{2\hbar^2m\gamma} +\frac{\gamma}{8mk_BT} \right) (p_2-p_1)^2 t \right]$. Using the fact that
\begin{equation}
\R_t(0,0)=\tr\ \rho_t =1\ ,
\end{equation}
we finally reach
\begin{equation}
\label{eq:thermal_density_matrix}
\bra{p_1} \rho_{\infty} \ket{p_2} = \left\{
\begin{array}{cl}
\sqrt{\frac{1}{2\pi mk_BT}}\ e^{-\frac{p_1^2}{2mk_BT}}&  p_1 = p_2\\
0 & \text{otherwise}
\end{array} \right.
\end{equation}
which is the stationary solution of the master equation and the thermal equilibrium density matrix of a free particle (see Sec.~(\ref{sec:setting_master})). Thus, the thermal equilibrium is indeed reached in the long time limit for the free particle in the Caldeira-Leggett model.


All of the asymptotic behaviour of the density matrix can be read off from~(\ref{eq:density_matrix_evolution}). Let us start by analyzing the remainder term $R(t)$. For large times and for the values of $x$ where the integrand is significantly different from $0$, we can series expand $\R_0$ through the leading term
\begin{align}
\R_0(p_2-p_1,xe^{-2\gamma t} + \frac{\Gamma_t (p_2-p_1)}{2m\gamma}) - \R_0(p_2-p_1, \frac{\Gamma_t (p_2-p_1)}{2m\gamma}) \approx \left. \frac{\partial \R_0(k,x)}{\partial x}\right|_{(p_2-p_1, \frac{ (p_2-p_1)}{2m\gamma})}\ xe^{-2\gamma t},
\end{align}
given that $\R_0$ is well-behaved around $x = (p_2-p_1)/2m\gamma$ and $\partial \R_0(k,x)/\partial x$ does not vanish at the point $ (p_2-p_1, \frac{ (p_2-p_1)}{2m\gamma})$. Under these conditions, we can calculate $R(t)$:
\begin{align}
R(t \to \infty) &\approx \frac{1}{2\pi \hbar} e^{-\left( \frac{k_BT}{2\hbar^2m\gamma} +\frac{\gamma}{8mk_BT} \right) (p_2-p_1)^2 t }\ e^{\frac{(p_2-p_1)^2k_BT}{8\hbar^2m\gamma^2} (\Gamma_t^2 +2\Gamma_t)}\ e^{-2\gamma t} \left. \frac{\partial \R_0(k,x)}{\partial x}\right|_{(p_2-p_1, \frac{ (p_2-p_1)}{2m\gamma})}\nonumber\\
& \hspace{20mm} \times \int dx\ x\ e^{-\frac{(1-e^{-4\gamma t})mk_BT}{2\hbar^2} x^2 -\left(\frac{i (p_1+p_2)}{2 \hbar} + \frac{ \Gamma_t^2k_B T}{2\hbar^2 \gamma} (p_2-p_1) \right) x} \nonumber\\
& = -\frac{\hbar^2}{\sqrt{2\pi((1-e^{-4\gamma t})mk_BT)^3}} \left(\frac{i (p_1+p_2)}{2 \hbar} + \frac{ \Gamma_t^2k_B T}{2\hbar^2 \gamma} (p_2-p_1) \right) \left. \frac{\partial \R_0(k,x)}{\partial x}\right|_{(p_2-p_1, \frac{ (p_2-p_1)}{2m\gamma})} \nonumber\\
& \hspace{20mm} \times e^{\frac{\hbar^2}{2(1-e^{-4\gamma t})mk_BT} \left(\frac{i (p_1+p_2)}{2 \hbar} + \frac{ \Gamma_t^2k_B T}{2\hbar^2 \gamma} (p_2-p_1) \right)^2}\  e^{\frac{(p_2-p_1)^2k_BT}{8\hbar^2m\gamma^2} (\Gamma_t^2 +2\Gamma_t)}\nonumber\\
&\hspace{30mm} \times e^{-\left( \frac{k_BT}{2\hbar^2m\gamma} +\frac{\gamma}{8mk_BT} \right) (p_2-p_1)^2 t }\ e^{-2\gamma t} \nonumber\\
& \approx f(p_1,p_2)\ e^{-\left( \frac{k_BT}{2\hbar^2m\gamma} +\frac{\gamma}{8mk_BT} \right) (p_2-p_1)^2 t }\ e^{-2\gamma t}\ . \nonumber\\
\label{eq:R_asymptotic}
\end{align}
where
\begin{align}
f(p_1,p_2)&=-\frac{\hbar^2}{\sqrt{2\pi(mk_BT)^3}} \left(\frac{i (p_1+p_2)}{2 \hbar} + \frac{k_B T}{2\hbar^2 \gamma} (p_2-p_1) \right) \left. \frac{\partial \R_0(k,x)}{\partial x}\right|_{(p_2-p_1, \frac{ (p_2-p_1)}{2m\gamma})} \nonumber\\
& \hspace{20mm} \times e^{\frac{\hbar^2}{2(1-e^{-4\gamma t})mk_BT} \left(\frac{i (p_1+p_2)}{2 \hbar} + \frac{ \Gamma_t^2k_B T}{2\hbar^2 \gamma} (p_2-p_1) \right)^2}\ e^{\frac{3(p_2-p_1)^2k_BT}{8\hbar^2m\gamma^2} }
\end{align}
This means, for the non-diagonal elements, $R(t)$ is dying much faster than the non-$R(t)$ term in~(\ref{eq:density_matrix_evolution}), due to the extra exponential factor of $e^{-2\gamma t}$, and is negligible. Then, the long time behaviour becomes
\begin{align}
\bra{p_1} \rho_{t \to \infty} \ket{p_2} &=
\sqrt{\frac{1}{2\pi m k_BT}}\ e^{\frac{(p_2-p_1)^2k_BT}{2\hbar^2m\gamma^2}}\  e^{- \frac{ (p_1+p_2)^2}{8 mk_BT}}\ e^{\frac{i(p_2^2-p_1^2)}{4m\hbar\gamma}}\ \R_0(p_2-p_1, \frac{ p_2-p_1}{2m\gamma}) \nonumber\\
& \hspace{20mm} \times e^{-\left( \frac{k_BT}{2\hbar^2m\gamma} +\frac{\gamma}{8mk_BT} \right) (p_2-p_1)^2 t } \qquad (p_1 \neq p_2)
\label{eq:free_matrix_elmnt_non_diagonal}
\end{align}
which is manifestly exponential. The time constant for relaxation is
\begin{align}
\label{eq:nondiag_time_const}
\tau_{p_1,p_2} = \left( \frac{\gamma}{8mk_BT} +\frac{k_BT}{2\hbar^2m\gamma} \right)^{-1} \frac{1}{(p_2-p_1)^2} = \frac{1}{1+ \left( \frac{\hbar \gamma}{2 k_B T} \right)^2}\ \frac{2\hbar^2m\gamma} {k_BT(p_2-p_1)^2}
\end{align}
which leads to the following observations:
\begin{enumerate}
\item In the momentum representation, relaxation to the thermal state becomes faster as one moves away from the diagonal in the density matrix.
\item At a first look, relaxation is faster in both high and low temperature limits  and is slowest at $T=\hbar\gamma/2k_B$, but remembering that $\hbar\gamma/k_BT \ll 1$ should hold for the validity of our master equation, only the high temperature limit is meaningful and the $\gamma/8mk_BT$ term can be neglected:
    \begin{align}
    \tau_{p_1,p_2} \approx \frac{2m\hbar^2\gamma} {k_BT(p_2-p_1)^2} =\frac{2\hbar^2}{D(p_2-p_1)^2}.
    \end{align}
\end{enumerate}
Here, $D = k_BT/m\gamma$ is the diffusion constant, which can be readily demonstrated using~(\ref{eq:density_matrix_evolution})
\begin{align}
\langle q^2 \rangle &= \tr (q^2 \rho_t) \nonumber\\
&=\int dp\ \bra{p} q^2 \rho_t \ket{p} \nonumber\\
&=-\hbar^2 \int dp \left( \frac{d^2}{dq^2} \bra{q} \rho_t \ket{p} \right)_{p=q} \nonumber\\
&=-\hbar^2 \int dp  \left( -2\left( \frac{k_BT}{2\hbar^2m\gamma} +\frac{\gamma}{8mk_BT} \right)t\ \bra{q} \rho_t \ket{p} + \textit{time independent terms} \right) \nonumber\\
& \approx \frac{k_BT}{m\gamma} t\ \ \ (t\to \infty)\ ,
\end{align}
where we again used the fact that the trace of the reduced density matrix is unity. This is the behaviour of the classical Brownian particle, and shows the connection between the decay constants and the diffusion coefficient.

For the diagonal elements, i.e. $p_1=p_2$, there are two sources of correction to~(\ref{eq:thermal_density_matrix}), one arising from $R(t)$, and the other from the corrections to the non-$R(t)$ term due to the finiteness of $t$. The latter is at the order $e^{-4\gamma t}$, as can be seen from~(\ref{eq:density_matrix_evolution}), so the leading correction comes from $R(t)$
\begin{align}
\bra{p} \rho_{t \to \infty} \ket{p}
&\approx \sqrt{\frac{1}{2\pi mk_BT}}\ e^{-\frac{p^2}{2mk_BT}} \left(1- \frac{ i \hbar p}{mk_BT} \left. \frac{\partial \R_0(k,x)}{\partial x}\right|_{(0,0)} e^{-2\gamma t} \right),
\end{align}

which shows that the relaxation time for the system is indeed $\tau_R \sim 1/\gamma$.


\section{The Time Evolution for the Non-Lindblad Master Equation}
\label{sec:non-Lindblad}

The originally derived master equation for the Caldeira-Leggett model, (\ref{eq:master_density_nonLindblad}), was not in the Lindblad form, lacking the ``minimally invasive term'' $-\frac{\gamma}{8mk_B T} [p,[p,\rho_t]]$.
The only difference not having this term makes in our previous analysis is that, when solving the differential equations for the coefficients of our Gaussian ansatz, the first equation reads
\begin{equation}
\dot{c}_1 (t) = \frac{c_2 (t)}{m}\ ,
\end{equation}
which is missing the $\gamma/8m k_BT$ term. All other differential equations for $c_i(t)$ are the same as before. This in turn leads to the formally simple change that $e^{-\frac{\gamma (p_2-p_1)^2}{8 m k_B}t}$ term is not present in~(\ref{eq:density_matrix_evolution}):
\begin{align}
\bra{p_1} \rho_t \ket{p_2} &= R(t) \nonumber\\
&+ \sqrt{\frac{1}{2\pi mk_BT}} \frac{1}{\sqrt{1-e^{-4\gamma t}}} e^{- \frac{k_BT}{2\hbar^2m\gamma} (p_2-p_1)^2 t }\ e^{\frac{(\Gamma_t^2 +2\Gamma_t)k_BT}{8\hbar^2m\gamma^2} (p_2-p_1)^2}\\ \nonumber\\
& \hspace{20mm} \times e^{\frac{\hbar^2}{2(1-e^{-4\gamma t})mk_BT} \left(\frac{i (p_1+p_2)}{2 \hbar} + \frac{ \Gamma_t^2k_B T}{2\hbar^2 \gamma} (p_2-p_1) \right)^2}\  \R_0(p_2-p_1, \frac{\Gamma_t (p_2-p_1)}{2m\gamma})
\end{align}

This means:
\begin{enumerate}
\item Any initial density matrix still evolves into the form in~(\ref{eq:thermal_density_matrix}) in the long time limit. So in the long time limit, the non-Lindblad and Lindblad equations give the same result, which is thermal equilibrium.
\item Non-diagonal terms still vanish in the long time limit due to the $e^{- \frac{k_BT}{2\hbar^2m\gamma} (p_2-p_1)^2 t }$ term, but the decay at first seems to be slower and the time constant is
    \begin{equation}
        \tau_{p_1,p_2} = \frac{2m\hbar^2\gamma} {k_BT(p_2-p_1)^2}
    \end{equation}
    as opposed to~(\ref{eq:nondiag_time_const}). Nevertheless, remembering the condition $\hbar \gamma \ll \min\{\hbar \Omega,2\pi k_BT\}$,
    we can see that the missing term in the non-Lindblad case is negligible, so the change in the time evolution for the non-diagonal terms is negligible.
\item Since the $e^{-\frac{\gamma (p_2-p_1)^2}{8 m k_B}t}$ term was constant and equal to $1$ for the diagonal matrix elements in the Lindblad form, not having this term does not have any effect. The time evolution of the diagonal elements of the density matrix is exactly the same as before.
\end{enumerate}

\section{Time evolution of the Simple Harmonic Oscillator System}
\label{sec:oscillator}
For a general potential
\begin{equation}
V(q) = \sum_m a_m q^{m},
\end{equation}
we cannot follow our method that solved the free particle case, since the equation contains terms of order higher than $2$, and in that case Gaussian solutions are not preserved. Still, it is possible to have an analytical solution for the exceptional, but important, case of the harmonic oscillator. For a system under the potential
\begin{equation}
V(q) = \frac{1}{2} m \omega^2 q^2\ ,
\end{equation}
all the calculational steps are very similar to, but algebraically more complicated than, those for the free particle, and they lead to (see  appendix~\ref{app:harmonic_greens})
\begin{multline} \label{eq:harmonic_greens}
   \tilde{G}(k,x,t;k_0,x_0,0)=\\
   \delta \left(k_0- \frac{e^{- (\gamma -\mu )t}\ \Lambda_t}{2
   \mu }\left( (\mu  \coth \mu t+ \gamma  )k- m  \omega ^2 x \right) \right)\ \delta \left(x_0- \frac{e^{- (\gamma -\mu )t}\ \Lambda_t}{2
   \mu }\left( ( \mu  \coth \mu t-\gamma) x +\frac{k}{m} \right) \right)\\ \times
\exp \left\{ -\frac{k_B T}{2 \hbar^2 m \omega^2} \left[M_1(t)k^2+\frac{2m \omega^2}{\gamma} M_2(t)k x +m^2\omega^2 M_3(t)x^2 \right] \right \}\ ,
\end{multline}
and
\begin{multline}
\R_t(k,x) = \exp \left\{ -\frac{  m k_BT M_3(t)}{2 \hbar^2}x^2 \right\}\ \exp \left\{- \frac{ k_B T}{2 \hbar^2 m \omega^2 } \left[M_1(t)k^2+\frac{2m \omega^2}{\gamma} M_2(t)k x \right] \right \} \\
\R_0 \left(\frac{e^{- (\gamma -\mu )t}\ \Lambda_t}{2\mu }\left( (\mu  \coth \mu t+ \gamma  )k- m  \omega ^2 x \right),\  \frac{e^{- (\gamma -\mu )t}\ \Lambda_t}{2\mu }\ \left( (\mu  \coth \mu t-\gamma) x +\frac{k}{m} \right) \right)\ ,
\end{multline}
where $\mu \equiv \sqrt{\gamma^2-\omega^2}$,  $\Lambda_t \equiv 1-e^{-2\mu t}$ and $M_i$ are dimensionless functions with the asymptotic behaviours
\begin{align}
M_1(t \to \infty) &= 1+ \left( \frac{\hbar \gamma}{2 k_B T}\right)^2 +\left( \frac{\hbar \omega}{4k_B T} \right)^2 +\ \Op (e^{-Re\{ 2\gamma - \mu\}t}) \nonumber\\
M_2(t \to \infty) &= -\frac{1}{8} \left( \frac{\hbar \gamma}{k_B T}\right)^2\ +\ \Op (e^{-Re\{ 2\gamma - \mu\}t})  \nonumber\\
M_3(t \to \infty) &= 1+\left( \frac{\hbar \omega}{4k_B T}\right)^2 +\ \Op (e^{-Re\{ 2\gamma - \mu\}t})\ .
\label{eq:M_is}
\end{align}
For the case of the harmonic oscillator, we will use the position representation where the thermal equilibrium density matrix is given by~(\ref{eq:density_matrix_thermal_HO}). Using the inversion formula~(\ref{eq:inversion_formulae}), we have
\begin{align}
\bra{q_1} \rho_t \ket{q_2} &=\frac{1}{2\pi \hbar} \int dk e^{-\frac{i}{\hbar} k \left(\frac{q_1+q_2}{2}\right) } \R_t(k,q_1-q_2) \nonumber\\
&=R_{HO}(t)+\frac{1}{2\pi \hbar} e^{-\frac{  m k_BT M_3(t)}{2 \hbar^2}(q_1-q_2)^2} \int dk e^{-\frac{ k_B T M_1(t)}{2 \hbar^2 m \omega^2}k^2 -\left(\frac{i(q_1+q_2)}{2\hbar}+ \frac{ k_B T M_2(t)(q_1-q_2)}{\hbar^2 \gamma} \right)k} \nonumber\\
&=R_{HO}(t)+ \sqrt{\frac{m \omega^2}{2 \pi k_B T M_1(t)}}\ e^{-\frac{  m k_BT M_3(t)}{2 \hbar^2}(q_1-q_2)^2}
e^{\frac{\hbar^2 m \omega^2}{2 k_B T M_1(t)} \left(\frac{i(q_1+q_2)}{2\hbar}+ \frac{ k_B T M_2(t)(q_1-q_2)}{\hbar^2 \gamma} \right)^2},
\label{eq:HO_exact_density_matrix}
\end{align}
with
\begin{multline}
R_{HO}(t) = \frac{1}{2\pi \hbar}\ e^{-\frac{m k_BT M_3(t)}{2 \hbar^2}(q_1-q_2)^2} \int dk\ e^{-\frac{ k_B T M_1(t)}{2 \hbar^2 m \omega^2} k^2 -\left(\frac{i(q_1+q_2)}{2\hbar}+ \frac{ k_B T M_2(t)(q_1-q_2)}{\hbar^2 \gamma} \right)k} \left( \R_0(k',x')-1 \right)
\end{multline}
where
\begin{align*}
k' &= \frac{e^{- (\gamma -\mu )t}\ \Lambda_t}{2\mu }\left( (\mu  \coth \mu t+ \gamma  )k- m  \omega ^2 (q_1-q_2) \right) \\
x' &= \frac{e^{- (\gamma -\mu )t}\ \Lambda_t}{2\mu }\ \left( (\mu  \coth \mu t-\gamma) (q_1-q_2) +\frac{k}{m} \right)\ ,
\end{align*}
and we again used the fact that $\R_t(0,0)=1$.

Assuming that $\R_0$ is well behaved around $(0,0)$,
\begin{equation}
\lim_{t \to \infty} R_{HO}(t) =0\ .
\end{equation}
This means, in the long time limit
\begin{multline}
\bra{q_1} \rho_{\infty} \ket{q_2} = \sqrt{\frac{ m\omega^2}{2\pi k_B T }\left[1+ \left( \frac{\hbar \gamma}{2 k_B T}\right)^2 +\left( \frac{\hbar \omega}{4k_B T} \right)^2 \right]^{-1}}\ e^{- \frac{m \omega^2 }{8k_BT} \left[1+ \left( \frac{\hbar \gamma}{2 k_B T}\right)^2 +\left( \frac{\hbar \omega}{4k_B T} \right)^2 \right]^{-1} (q_1+q_2)^2}\\
\times e^{- \frac{mk_B T}{2\hbar^2}\ \left[ 1+ \left(\frac{\hbar \omega}{4k_B T} \right)^2 + \left(\frac{\hbar \omega}{4k_B T} \right)^2 \left(\frac{\hbar \gamma}{2k_B T} \right)^2 \left[ 1+ \left(\frac{\hbar \gamma}{2k_BT}\right)^2 +\left( \frac{\hbar \omega}{4k_B T} \right)^2 \right]^{-1} \right]  (q_1-q_2)^2}\\
\times e^{- i \frac{m \omega^2 }{16k_BT}\ \left( \frac{\hbar \gamma}{k_B T}\right) \left[1+ \left( \frac{\hbar \gamma}{2 k_B T}\right)^2 +\left( \frac{\hbar \omega}{4k_B T} \right)^2 \right]^{-1}(q_1^2-q_2^2)}
\label{eq:HO_apprx_stationary}
\end{multline}

This result does not exactly agree with the thermal equilibrium density matrix~(\ref{eq:density_matrix_thermal_HO}) or with the stationary solution to the non-Lindblad master equation~(\ref{eq:stationary_solution_general}). Nevertheless, conditions for the validity of the master equation imply that $ \hbar \gamma/k_B T \ll 1$ and $\hbar \omega/k_B T \ll 1$. Also, $m\omega^2 q^2 \sim \hbar \omega$ and
\begin{equation}
e^{- i \frac{m \omega^2 }{16k_BT}\ \left( \frac{\hbar \gamma}{k_B T}\right) (q_1^2-q_2^2)} \approx e^{- i \left( \frac{\hbar \omega}{16 k_B T}\right)\ \left( \frac{\hbar \gamma}{k_B T}\right) } \approx 1
\end{equation}
for the typical length scales we encounter in the harmonic oscillator system.
Putting all these conditions together:
\begin{equation}
\bra{q_1} \rho_{\infty} \ket{q_2} \approx \sqrt{\frac{ m\omega^2}{2\pi k_B T }}\ e^{- \frac{m \omega^2 }{8k_BT}  (q_1+q_2)^2}
e^{- \frac{mk_B T}{2\hbar^2} (q_1-q_2)^2}\ .
\end{equation}
So, the density matrix approaches the stationary solution~(\ref{eq:stationary_solution_general}) approximately, which is the expected result since~(\ref{eq:stationary_solution_general}) was derived for the non-Lindblad master equation. Also, remember that the stationary solution agrees with the thermal equilibrium density matrix of the harmonic oscillator only on the diagonal and to the leading order in $\hbar \omega/k_BT$.

Let us now analyze the corrections to the infinite time matrix elements. Assuming $\R_0$ is well behaved around $(0,0)$ and defining $\left. \partial \R_0(k,x)/\partial k \right|_{(0,0)} \equiv D_k$ and $\left. \partial \R_0(k,x)/\partial x \right|_{(0,0)} \equiv D_x$, the remainder term has the long time behaviour
\begin{align} \label{eq:harmonic_R_H}
R_{HO}(t) &\approx \frac{1}{2\pi \hbar} e^{-\frac{  m k_BT M_3(t)}{2 \hbar^2 }(q_1-q_2)^2}\ \frac{e^{- (\gamma -\mu )t}\ \Lambda_t}{2\mu } \left( (\mu \coth \mu t +\gamma)D_k + \frac{D_x}{m} \right) \nonumber\\
& \times \int dk e^{-\frac{ k_B T M_1(t)}{2 \hbar^2 m \omega^2}k^2-\left(\frac{i(q_1+q_2)}{2\hbar}+ \frac{ k_B T M_2(t)(q_1-q_2)}{\hbar^2 \gamma} \right)k}  \left( k+\frac{-m\omega^2 D_k +(\mu \coth \mu t -\gamma)D_x}{ (\mu \coth \mu t +\gamma)D_k + \frac{D_x}{m}} (q_1-q_2)\right)  \nonumber\\
& \approx  \sqrt{\frac{m \omega^2}{2 \pi k_B T M_1(t)}}\ e^{-\frac{  m k_BT M_3(t)}{2 \hbar^2}(q_1-q_2)^2}
e^{\frac{\hbar^2 m \omega^2}{2 k_B T M_1(t)} \left(\frac{i(q_1+q_2)}{2\hbar}+ \frac{ k_B T M_2(t)(q_1-q_2)}{\hbar^2 \gamma} \right)^2} \nonumber\\
&\hspace{1cm} \times \left( (\mu \coth \mu t +\gamma)D_k + (D_x/m)\right) \frac{ \Lambda_t}{2\mu }\ \nonumber\\
&\hspace{1cm} \times  \left( -\frac{\hbar m \omega^2 \left( 4ik_BT (q_1+q_2)-\hbar\gamma (q_1-q_2) \right)}{8k_B^2T^2} + \frac{-m\omega^2 D_k +(\mu \coth \mu t -\gamma)D_x}{ (\mu \coth \mu t +\gamma)D_k + (D_x/m)} (q_1-q_2) \right) \nonumber\\
& \hspace{1cm} \times e^{- (\gamma -\mu )t}
\end{align}
In the long time limit, again there are two sources of correction to the infinite time values of the matrix elements. The first is from $R_{HO}$ which is exponentially small compared to the matrix element at the infinite time limit  by $e^{-Re\{ \gamma - \mu\}t}$, as seen in~(\ref{eq:harmonic_R_H}). The second correction comes from the corrections to the non-$R_{HO}$ term due to the finiteness of $t$ in~(\ref{eq:HO_exact_density_matrix}), which mainly arises from $M_i(t) -M_i(\infty)$, and is at the order $ e^{-Re\{ 2\gamma - \mu\}t}$ as seen in~(\ref{eq:M_is}). Thus, after a sufficiently long time, the latter correction becomes negligible compared to the former unless $R_{HO}$ vanishes, which is only possible for a special combination of values of $D_k$ and $D_x$. That is, the leading correction to~(\ref{eq:HO_apprx_stationary}) is~(\ref{eq:harmonic_R_H}), which dies out with a time constant of $1/Re\{ \gamma - \mu\}$

If we were to use the non-Lindblad master equation, than the only change would be that we would have no $\gamma/8mk_BT$ term for $\dot{c}_1$ in~(\ref{eq:diff_eqn_SHO}). Effects of this can be easily traced by having $k_B T \to \infty$ and $\hbar \rightarrow \infty$ while keeping $\hbar^2/k_BT$ constant. This way, $\gamma/8mk_BT \rightarrow 0$, with all other coefficients in the differential equations remaining the same, giving us the non-Lindblad equation. Then
\begin{align}
\frac{\hbar \omega}{k_B T} &= \frac{\hbar^2}{k_B T} \frac{\omega}{\hbar} \rightarrow 0 \nonumber\\
\frac{\hbar \gamma}{k_B T} &= \frac{\hbar^2}{k_B T} \frac{\gamma}{\hbar} \rightarrow 0
\end{align}
In this limit, terms of $\Op (\hbar \gamma/k_B T)$ or $\Op (\hbar \omega/k_B T)$ in the expressions for $M_i$,~(\ref{eq:M_is}), vanish. Thus, without any approximations
\begin{equation}
\bra{q_1} \rho_{\infty} \ket{q_2} = \sqrt{\frac{ m\omega^2}{2\pi k_B T }}\ e^{- \frac{m \omega^2 }{8k_BT}  (q_1+q_2)^2}
e^{- \frac{mk_B T}{2\hbar^2} (q_1-q_2)^2}
\end{equation}
as expected. This result was also reached in~\cite{ref:exact_solution3}, where they also show that the density matrix is diagonal in the energy basis of the system harmonic oscillator, supporting the idea of the pointer states of Paz and Zurek~\cite{ref:pointer_states}.

For the non-Lindblad master equation, The leading correction to the thermal equilibrium density matrix at large times is the same as the Lindblad case and originates from $R_{HO}(t)$. Thus it is $\Op (e^{-Re\{ \gamma - \mu\}t})$. In short, non-Lindblad time evolution is equivalent to the Lindblad case within the sensitivity of our calculations.

Note the following observations for the harmonic oscillator system
\begin{enumerate}
    \item If $\omega > \gamma$, $\mu$ is imaginary, and the relaxation time constant is $1/Re\{ \gamma - \mu\} = 1/\gamma$.
    \item If $\omega < \gamma$, $\mu$ is real and $\mu < \gamma$, so relaxation still occurs but at the slower rate of $e^{-(\gamma-\sqrt{\gamma^2 -\omega^2})t}$. The rate decreases as $\frac{\omega}{\gamma}$ approaches $0$.
\end{enumerate}

\section{Caldeira-Leggett Model in Higher Dimensions} \label{sec:higher_dimensions}
For a quadratic potential and a generalized dot product interaction term of $-\sum_{i=1}^d q_i \sum_n \kappa_{n,i} q_{n,i}$, the Caldeira-Leggett Hamiltonian for a free particle or a harmonic oscillator in $d$-dimensions is separable
\begin{equation}
H = \sum_{i=1}^d \left\{\frac{1}{2m}p_i^2 + \frac{1}{2}m\omega_i^2 q_i^2+\  q_i^2 \sum_n \frac{\kappa_{n,i}^2}{2m_n\omega_{n,i}^2} +\ \sum_n \left( \frac{1}{2m_n}p_{n,i}^2+\frac{1}{2}m_n \omega_{n,i}^2q_{n,i}^2\right) -\ q_i \sum_n \kappa_{n,i} q_{n,i} \right\}\ .
\end{equation}
This means, we can write $\rho_t$ as a linear superposition of terms like $\rho^{(1)}_t \otimes...\rho^{(i)}_t...\otimes \rho^{(d)}_t $  where
\begin{align}
\frac{d}{dt}\rho^{(i)}_t &= -\frac{i}{\hbar}[H_{S,i},\rho^{(i)}_t] -\frac{2 \gamma_i mk_B T}{\hbar^2}[q_i,[q_i,\rho^{(i)}_t]] -\frac{\gamma_i}{8mk_B T} [p_i,[p_i,\rho^{(i)}_t]] -i \frac{\gamma_i}{\hbar} [q_i,\{p_i,\rho^{(i)}_t \}]\ .
\end{align}
We can solve for each dimension separately using its own parameter $\gamma_i$, take the tensor product and finally superpose such solutions to find the overall density matrix. So our results for the one dimensional case readily generalize to any higher dimensions.

\section{Conclusions} \label{sec:conclusions}

We have solved the complete time evolution of the Caldeira-Leggett model for the free particle and the harmonic oscillator, and showed that the reduced density matrix of the system approaches the exact thermal equilibrium for the free particle and an approximate thermal equilibrium for the harmonic oscillator in the long time limit. Moreover, we studied the leading corrections to the infinite time behaviour of the matrix elements of the density matrix at finite times, and calculated the time scales for the decay of these corrections. We discussed the deviations from the thermal equilibrium for the harmonic oscillator in the infinite time limit and showed that they follow from the approximations we introduced during the derivation of the master equations.

Our calculations explicitly showed that the typical time scale for reaching the thermal equilibrium is $\sim 1/\gamma$. Also we demonstrated that, within the limits of our master equations~(\ref{eq:master_density_nonLindblad}) and (\ref{eq:master_density}), the Lindblad and non-Lindblad forms give the same time evolution.

We note that the methods we used, the characteristic function of the Wigner function and the Gaussian ansatz, can be generalized to any equation of type~(\ref{eq:master_wigner}) as long as they do not lead to nonlinear variations of~(\ref{eq:SHO_c_i}) (see appendix~\ref{app:Wigner_master}). We gave a treatment of the most general equation in the appendices.

\section*{Acknowledgements}
We thank S.L.~Adler for his guidance throughout this project and comments on the early drafts of this article. We learned the method of the Gaussian ansatz for solving the Green's functions from A.~Bassi, and also thank him for valuable discussions and comments on the manuscript. The relation between the diffusion constant and the exponential decay of the matrix elements in the free particle case was pointed to us by D. Huse.
\appendix

\section{The Master Equation for the Characteristic Function associated with the Wigner Function for the Free Particle Case}
\label{app:Wigner_master}
In the position representation, $\rho_t(q_1,q_2) = \bra{q_1} \rho_t \ket{q_2}$,~(\ref{eq:master_density}) is
\begin{multline}
\frac{\partial \rho_t(q_1,q_2)}{\partial t} = \bigg[ \frac{i\hbar}{2m} \left( \frac{\partial^2}{\partial q_1^2} -\frac{\partial^2}{\partial q_2^2} \right) - \frac{i}{\hbar} (V(q_1)-V(q_2)) - \frac{2\gamma mk_BT}{\hbar^2} (q_1-q_2)^2 \\
+\frac{ \hbar^2 \gamma }{8mk_BT} \left(\frac{\partial}{\partial q_1} -\frac{\partial}{\partial q_2} \right)^2 - \gamma (q_1-q_2) \left( \frac{\partial}{\partial q_1} -\frac{\partial}{\partial q_2} \right) \bigg] \rho_t(q_1,q_2)
\label{eq:master_q_space}
\end{multline}
Since this equation is second order in $q_1, q_2$ and their derivatives, we can propose a Gaussian ansatz, $\rho_t(q_1,q_2) = \exp \{ -c_1q_1^2-c_2q_1q_2-c_3q_2^2-ic_4q_1-ic_5q_2 -c_6\}$, without using the characteristic function associated with the Wigner function. This will lead to a set of coupled differential equations as in~(\ref{eq:SHO_c_i}) which is in principle solvable, but the basic difference from~(\ref{eq:SHO_c_i}) is that these equations will be non-linear due to the double derivative terms, e.g.
\begin{align}
\frac{\partial^2}{\partial q_1^2} \rho_t(q_1,q_2)
&= \left[ (-2c_1q_1-c_2q_2-ic_4)^2-2c_1\right] \rho_t(q_1,q_2)
\end{align}
Thus they will be much more cumbersome to solve.

In  short, any equation of the form~(\ref{eq:master_q_space}) that contains double derivatives leads to a non-linear system of differential equations for $c_i$ when we employ a Gaussian ansatz. This is the basic reason for using the characteristic function associated with the Wigner function to solve the evolution problem, where we solve a linear system of differential equations for $c_i$. See appendix~\ref{app:general_treatment} for more discussions.

Let us now demonstrate how~(\ref{eq:master_wigner}) arises. The following identity follows from the Baker-Campbell-Hausdorff formula and will be useful in our calculations:
\begin{align}
e^{\frac{i}{\hbar}(kq+xp)} &= e^{\frac{i}{\hbar}kq}\ e^{\frac{i}{\hbar}xp}\ e^{\frac{i}{\hbar}\frac{kx}{2}} = e^{\frac{i}{\hbar}xp}\ e^{\frac{i}{\hbar}kq}\ e^{-\frac{i}{\hbar}\frac{kx}{2}}.
\end{align}

Equation~(\ref{eq:wigner_definition}) is derived as
\begin{align}
\tilde{\rho}_t(k,x) &= \tr \left( \rho_t e^{\frac{i}{\hbar}(kq+xp)} \right) \nonumber\\
&= \int du'\ \bra{u'} \rho_t e^{\frac{i}{\hbar}(kq+xp)} \ket{u'} \nonumber\\
&= \int du'\ \int dv\ \bra{u'} \rho_t \ket{v} \bra{v} e^{\frac{i}{\hbar}kq}\ e^{\frac{i}{\hbar}xp}\ e^{\frac{i}{2 \hbar}kx} \ket{u'} \nonumber\\
&= \int du'\ e^{\frac{i}{\hbar}x \left( u'+\frac{k}{2}\right)}\ \int dv\ \bra{u'} \rho_t \ket{v} \langle v|u'+k \rangle \nonumber\\
&= \int du\ e^{\frac{i}{\hbar}xu} \bra{u-\frac{k}{2}} \rho_t \ket{u+\frac{k}{2}}\ ,
\end{align}
$\{\ket{u'}\}$ and $\{\ket{v}\}$ representing the momentum basis, and the inversion formula is
\begin{align}
\bra{p_1} \rho_t \ket{p_2} &=\int du\ \delta(u-\frac{p_1+p_2}{2}) \bra{u-\frac{p_2-p_1}{2}} \rho_t \ket{u+\frac{p_2-p_1}{2}} \nonumber\\
&= \int du\ \frac{1}{2\pi \hbar} \int dx\ e^{\frac{i}{\hbar} x(u-\frac{p_1+p_2}{2})} \bra{u-\frac{p_2-p_1}{2}} \rho_t \ket{u+\frac{p_2-p_1}{2}} \nonumber\\
&= \frac{1}{2\pi \hbar} \int dx\ e^{-\frac{i}{\hbar} x(\frac{p_1+p_2}{2})} \int du\ e^{\frac{i}{\hbar} xu} \bra{u-\frac{p_2-p_1}{2}} \rho_t \ket{u+\frac{p_2-p_1}{2}} \nonumber\\
&= \frac{1}{2\pi \hbar} \int dx\ e^{-\frac{i}{\hbar} x(\frac{p_1+p_2}{2})}\ \tilde{\rho}_t (p_2-p_1,x)\ .
\end{align}
The definition and the inversion formula for the position basis are derived similarly.

To obtain~(\ref{eq:master_wigner}) from~(\ref{eq:master_density}), we will need the following identities which hold for operators $\Op$ and $\V$ and which follow from the basic commutation relations and the cyclicity of the trace:
\begin{align}
\tr \left( [\V,\Op] e^{\frac{i}{\hbar}(kq+xp)} \right) &=\tr \left( \Op [e^{\frac{i}{\hbar}(kq+xp) },\V] \right)\\
[f(\V),\Op] &= cf'(\V) \hspace{10mm} \text{if} \ [\V,\Op]=c \ \text{a constant}\ .
\end{align}

To calculate the first term on the right hand side of~(\ref{eq:master_wigner}), note that
\begin{align}
\frac{\partial}{\partial x} e^{\frac{i}{\hbar}(kq+xp)} &=\frac{\partial}{\partial x} \left( e^{\frac{i}{\hbar}kq}\ e^{\frac{i}{\hbar}xp}\ e^{\frac{i}{\hbar}\frac{kx}{2}} \right)= \frac{i}{\hbar}\ e^{\frac{i}{\hbar}(kq+xp)}\ (p+\frac{k}{2})\nonumber\\
&=\frac{\partial}{\partial x} \left( e^{\frac{i}{\hbar}xp}\ e^{\frac{i}{\hbar}kq}\ e^{- \frac{i}{\hbar}\frac{kx}{2}} \right)= \frac{i}{\hbar} (p-\frac{k}{2}) e^{\frac{i}{\hbar}(kq+xp)}\ \nonumber\\
=& \frac{i}{2\hbar} \left( p\ e^{\frac{i}{\hbar}(kq+xp)}+e^{\frac{i}{\hbar}(kq+xp)}\ p \right).
\end{align}
which in turn implies
\begin{align}
\tr \left( \{p, \Op \}e^{\frac{i}{\hbar}(kq+xp)} \right) = -2i \hbar \frac{\partial}{\partial x} \tr \left( \Op e^{\frac{i}{\hbar}(kq+xp)} \right)
\end{align}
for any $x$-independent operator $\Op$. This gives
\begin{align}
\tr \left( [\frac{1}{2m} p^2,\rho_t] e^{\frac{i}{\hbar}(kq+xp)}\right) &=\frac{1}{2m}\ \tr \left( \rho_t \{p, [e^{\frac{i}{\hbar}(kq+xp)},p] \}\right)\nonumber\\
&= -\frac{1}{2m}\ k\ \tr \left( \rho_t \{p,e^{\frac{i}{\hbar}(kq+xp)} \}\right) \nonumber\\
&=\frac{i \hbar }{m}\ k \frac{\partial}{\partial x}\tr \left( \rho_t e^{\frac{i}{\hbar}(kq+xp)} \right)= \frac{i \hbar }{m}\ k \frac{\partial}{\partial x} \R_t(k,x)
\label{eq:kinetic_energy_to_wigner}
\end{align}

For the last term on the right hand side of~(\ref{eq:master_wigner}),
\begin{align}
\tr \left( [q,\{p,\rho_t\}] e^{\frac{i}{\hbar}(kq+xp)}\right) &= \tr \left( \rho_t \{p, [e^{\frac{i}{\hbar}(kq+xp)},q] \}\right) \nonumber\\
&= x\ \tr \left( \rho_t \{p,e^{\frac{i}{\hbar}(kq+xp)} \}\right) \nonumber\\
&=-2 i \hbar\ x \frac{\partial}{\partial x}\tr \left( \rho_t e^{\frac{i}{\hbar}(kq+xp)} \right)= -2 i \hbar\ x \frac{\partial}{\partial x} \R_t(k,x)\ .
\end{align}

The second and the third terms are calculated in a similar way:
\begin{align}
\tr \left( [q,[q,\rho_t]] e^{\frac{i}{\hbar}(kq+xp)}\right) &= \tr \left( \rho_t [ [e^{\frac{i}{\hbar}(kq+xp)},q],q ]\right) = x^2 \R_t(k,x)\\
\tr \left( [p,[p,\rho_t]] e^{\frac{i}{\hbar}(kq+xp)}\right) &= \tr \left( \rho_t [ [e^{\frac{i}{\hbar}(kq+xp)},p],p ]\right) = k^2 \R_t(k,x)\ .
\end{align}

Putting all the terms together,~(\ref{eq:master_wigner}) is obtained.

\section{Green's Function for the Harmonic Oscillator System}
\label{app:harmonic_greens}
First, we calculate the contribution of the harmonic potential term to the differential equation for $\R_t(k,x)$ in a way similar to~(\ref{eq:kinetic_energy_to_wigner}):
\begin{align}
\tr \left( [\frac{1}{2} m \omega^2 q^2,\rho_t] e^{\frac{i}{\hbar}(kq+xp)}\right) &=\frac{1}{2}m\omega^2 \ \tr \left( \rho_t \{q, [e^{\frac{i}{\hbar}(kq+xp)},q] \}\right)\nonumber\\
&= \frac{1}{2}m\omega^2x\ \tr \left( \rho_t \{q,e^{\frac{i}{\hbar}(kq+xp)} \}\right) \nonumber\\
&=-i\hbar m\omega^2\ x \frac{\partial}{\partial k}\ \tr \left( \rho_t e^{\frac{i}{\hbar}(kq+xp)} \right)= -i\hbar m\omega^2\ x \frac{\partial}{\partial k} \R_t(k,x),
\end{align}
Together with this term, the equations for the coefficients in our Gaussian ansatz become:
\begin{align}
\begin{split}
   \dot{c}_1 (t) &= \frac{c_2 (t)}{m}+ \frac{\gamma}{8mk_B T} \\
   \dot{c}_2 (t) &=  \frac{2c_3 (t)}{m}-2\gamma c_2 (t)-2m\omega^2c_1(t)\\
   \dot{c}_3 (t) &= \frac{2\gamma mk_BT}{\hbar^2}-4\gamma  c_3 (t)-m\omega^2 c_2(t)\\
   \dot{c}_4 (t) &=  \frac{c_5 (t)}{m}\\
   \dot{c}_5 (t) &=  -2\gamma c_5 (t) -m\omega^2 c_4(t)\\
   \dot{c}_6 (t) &=0 \label{eq:diff_eqn_SHO}
\end{split}
\end{align}
This system of equations is still linear but much more cumbersome to solve by hand. Hence, we used Mathematica$^{\copyright}$ to solve it for the initial value
\begin{equation}
   \tilde{\rho}_0^{k_0x_0,\epsilon\eta} (k,x)
   =\frac{1}{\pi\sqrt{\epsilon\eta}} e^{-\frac{1}{\epsilon} (k-k_0)^2} e^{-\frac{1}{\eta} (x-x_0)^2}\  .
\end{equation}
We will not give the details of the calculation, which are quite lengthy. Upon rearranging the terms into complete squares and defining $\mu \equiv \sqrt{\gamma^2-\omega^2}$,  $\Lambda_t \equiv 1-e^{-2\mu t}$, we reach
\begin{multline}
   \tilde{\rho}_t^{k_0x_0,\epsilon\eta}
   (k,x)=\frac{1}{\pi\sqrt{\epsilon\eta}}\ e^{-\frac{1}{\epsilon}
     \left[k_0- \frac{e^{- (\gamma -\mu )t}\ \Lambda_t}{2
   \mu }\left((\mu  \coth \mu t+ \gamma  )k- m  \omega ^2 x
    \right) \right]^2}\ \\
\times e^{-\frac{1}{\eta} \left[x_0- \frac{e^{- (\gamma -\mu )t}\ \Lambda_t}{2
   \mu }\left( ( \mu  \coth \mu t-\gamma) x +\frac{k}{m} \right) \right]^2}\\ \times
e^{-\frac{ k_B T}{2 \hbar^2 m \omega^2} \left[M_1(t)k^2+\frac{2m \omega^2}{\gamma} M_2(t)k x +m^2\omega^2 M_3(t)x^2 \right]}
\end{multline}
where
\begin{align}
M_1(t) &= -\frac{1}{\mu^2} \left[(e^{-2\gamma t}\ \cosh 2\mu t - 1)\ \gamma^2+ \Gamma_t\ \omega^2+ e^{-2\gamma t}\ \sinh 2\mu t\ \gamma\mu \right]\nonumber\\
-&\frac{\hbar^2}{16k_B^2T^2\mu^2} \left[4 (e^{-2\gamma t}\ \cosh 2\mu t -1)\ \gamma^4 -3 (e^{-2\gamma t}\ \cosh 2\mu t -1)\ \gamma^2 \omega^2 \right. \nonumber\\
& \hspace{20mm} +\Gamma_t\ \omega^4 + 4 e^{-2\gamma t}\ \sinh 2\mu t\ \gamma^3 \mu -e^{-2\gamma t}\  \sinh 2\mu t\ \gamma \omega^2 \mu \left. \right] \nonumber\\
M_2(t) &= \frac{\gamma^2}{2\mu^2} e^{-2(\gamma - \mu) t}\ \Lambda_t^2 \nonumber\\
&+\frac{\hbar^2 \gamma^2}{16k_B^2T^2 \mu^2} \left[2(e^{-2\gamma t}\ \cosh 2\mu t-1) \gamma^2 -(e^{-2\gamma t}+ e^{-2\gamma t}\ \cosh 2\mu t-2)\ \omega^2 + 2e^{-2\gamma t}\ \sinh 2\mu t\ \gamma \mu \right]
\end{align}
\begin{align}
M_3(t) &= \frac{1}{\mu^2} \left[-(e^{-2\gamma t}\ \cosh 2\mu t - 1)\ \gamma^2- \Gamma_t\ \omega^2+ e^{-2\gamma t}\ \sinh 2\mu t\ \gamma\mu \right]\nonumber\\
& \hspace{10mm} -\frac{\hbar^2\omega^2}{16k_B^2T^2\mu^2} \left[(e^{-2\gamma t}\ \cosh 2\mu t - 1)\ \gamma^2+ \Gamma_t\ \omega^2+ e^{-2\gamma t}\ \sinh 2\mu t\ \gamma\mu \right]\ .
\end{align}
%


Taking the limits $\epsilon \to 0$ and $\eta \to 0$, we obtain~(\ref{eq:harmonic_greens}).

\section{The Free particle as the Limit of the Harmonic Oscillator}

Note that the harmonic oscillator hamiltonian gives the free particle hamiltonian in the $\omega \rightarrow 0$ limit, so we expect the Green's function and the density matrix of the harmonic oscillator to converge to those of the free particle in this limit. Thus, one can obtain the results for the free particle by first solving the problem for the harmonic oscillator and then taking the said limit. In our strategy, we will rather use this correspondence as an independent check of our results. Using Taylor series, we can expand $M_i(t)$ around $\omega=0$:
\begin{align}
M_1(t) &= \frac{t \gamma  \left(h^2 \beta ^2 \gamma ^2+4\right)-(3-4 e^{-2 t \gamma }+e^{-4 t \gamma} ) }{4\gamma^2}\ \omega^2+ \Op(\omega^4) \nonumber \\
M_2(t) &= \frac{1}{2} \Gamma_t^2 +\Op(\omega^2) \\
M_3(t) &= (1-e^{-4\gamma t}) + \Op(\omega^2). \nonumber
\end{align}
Note also that $\mu \rightarrow \gamma$ in the vanishing $\omega$ limit. Then, it is trivial to recover~(\ref{eq:free_greens_function}) by inserting the above expressions into~(\ref{eq:harmonic_greens}) and taking the $\omega \rightarrow 0$ limit.

%
%

\section{The most General Case for the Gaussian Ansatz}\label{app:general_treatment}
In this appendix, we discuss the most general equation for which the Gaussian ansatz can be employed.

Gaussian ansatz is applicable to any equation of the form
\begin{align}
\frac{\partial f_t(k,x)}{\partial t}
&= \bigg[A \nonumber \\
&+B\ k+C\ x \nonumber\\
&+D\ \frac{\partial}{\partial k} +E\ \frac{\partial}{\partial x} \nonumber\\
&+F\ k^2+G\ kx+H\ x^2 \nonumber\\
&+L\ k \frac{\partial}{\partial k} +M\ k \frac{\partial}{\partial x} + N\ x \frac{\partial}{\partial k} + P\ x  \frac{\partial}{\partial x} \nonumber\\
&+Q\  \frac{\partial^2}{\partial k^2} +R\  \frac{\partial^2}{\partial k \partial x} +S\  \frac{\partial^2}{\partial x^2} \bigg] f_t(k,x)
\label{eq:most_general_partial}
\end{align}

We argued in appendix~\ref{app:Wigner_master} that we have to solve a nonlinear system of differential equation unless $Q,R,S=0$. One special case we can avoid nonlinearity is when $F,G,H=0$. In that case, we can Fourier transform $f$ in both $k$ and $x$, and since the Fourier transform converts differentiation into multiplication, we do not have second order derivatives in the transformed equation. A single Fourier transformation can also be useful when $F,R,S=0$ and $Q\neq 0$, or $H,Q,R=0$ and $S\neq 0$. Roy and Venugopalan successfully use this approach in~\cite{ref:exact_solution3} to solve the time evolution of the harmonic oscillator density matrix for the non-Lindblad master equation, after certain change of coordinates in the position representation. However, if the ``minimally invasive term'' is introduced (which they do not attempt to do), their method cannot avoid having a second order derivative. For the rest of our discussion, we will set $Q,R,S =0$ and use the shorthand notation
\begin{equation}
\frac{\partial f_t(k,x)}{\partial t}=\mathcal{D}(A,B,C,D,E,F,G,H,L,M,N,P) f_t(k,x)
\end{equation}

When we propose a Gaussian ansatz of the form~(\ref{eq:gaussian_ansatz}), we reach the following system of coupled linear equations

\begin{equation}
\left( \begin{array}{c}
   \dot{c}_1 (t) \\
   \dot{c}_2 (t)\\
   \dot{c}_3 (t)\\
   \dot{c}_4 (t)\\
   \dot{c}_5 (t)\\
   \dot{c}_6 (t)
\end{array} \right)
 =\left( \begin{array}{cccccc}
2L &M &0 &0 &0&0 \\
2N &L+P &2M &0 &0&0\\
 0&N    &2P  &0 &0&0\\
-2iD &-iE &0 &L &M&0\\
 0&-iD &-2iE &N &P&0\\
 0&0&0&iD&iE&0
\end{array} \right)
\left( \begin{array}{c}
c_1\\c_2\\c_3\\c_4\\c_5\\c_6 \end{array} \right)
+
\left( \begin{array}{c}
-F\\-G\\-H\\iB\\iC\\-A \end{array} \right)
\label{eq:most_general_c_i}
\end{equation}

This is a system of inhomogeneous ordinary linear differential equations which can be solved by basic methods, but the dimension of the matrix makes the solution intractable from a calculational point of view, even for mathematical software packages.

The first observation that makes the calculation considerably easier is that $c_1,c_2,c_3$ form an independent system of equations. This means, we can first solve for these three, then insert the solutions into the equations for $c_4$ and $c_5$ and solve the inhomogeneous equations for these two variables. We can finally insert $c_4, c_5$ into the equation for $c_6$ and find the solution by simple integration. This approach is tractable for Mathematica$^{\copyright}$, but the solutions are rather lengthy and give us little insight.

The crucial step that simplifies~(\ref{eq:most_general_c_i}) is that by an affine transformation of the variables $k,x$ in~(\ref{eq:most_general_partial}), we can set the coupling terms $D,E,M,N$ to $0$ for most cases, and have a diagonal matrix in~(\ref{eq:most_general_c_i}). Let us define the variables $l, y$ such that
\begin{align}
k &=  l\ +ay \nonumber\\
x &= bl\ + y\ ,
\end{align}
which together with the scaling and swapping ($k \leftrightarrow x$) can account for all linear transformations. This leads to the equation
\begin{equation}
\frac{\partial f^{(ly)}_t(l,y)}{\partial t}=\mathcal{D}(A',B',C',D',E',F',G',H',L',M',N',P') f^{(ly)}_t(l,y)
\end{equation}
with
\begin{align}
M' &=\frac{1}{1-ab} \left( -bL+M-b^2N+bP \right) \nonumber\\
N' &=\frac{1}{1-ab} \left( aL-a^2 M+N-aP \right).
\end{align}
By choosing
\begin{align}
a &=\frac{(L-P)+\sqrt{(L-P)^2+4MN}}{2M} \nonumber\\
b &=\frac{-(L-P)-\sqrt{(L-P)^2+4MN}}{2N}\
\end{align}
if $M$ and $N$ are both nonzero, and
\begin{align}
a &=0 \nonumber\\
b &=\frac{M}{L-P}\
\end{align}
if $N=0$ (the case of $M=0$ is similar), we can set $M'$ and $N'$ to $0$. The signs of the roots of the quadratics are chosen such that $ab \neq 1$, which ensures the linear independence of $l$ and $y$. Note that this procedure cannot be used if $MN=0$ and $L=P$.

Once we set $M',N'=0$, given that $L'$ and $P'$ are nonzero, we can shift our variables as
\begin{align}
m &\equiv l+\frac{D'}{L'} \nonumber\\
z &\equiv y+\frac{E'}{P'}\ ,
\end{align}
which puts our equation into the form
\begin{equation}
\frac{\partial f^{(mz)}_t(m,z)}{\partial t}=\mathcal{D}(A'',B'',C'',0,0,F',G',H',L',0,0,P') f^{(mz)}_t(m,z)\ .
\end{equation}

This equation leads to six inhomogeneous ordinary differential equations which are not coupled, and thus can be solved quite easily. One can further simplify the equations if $F' \neq0$, by scaling $k \rightarrow \sqrt{F'}k$ to set the coefficient of the $k^2$ term to $1$. By defining the function $\tilde{f}^{(mz)}_t=f^{(mz)}_te^{A''t}$, the constant term $A''$ can also be set to 0.

The special cases we did not discuss, e.g. $MN =0$ and $P=L$, can also be handled using similar techniques. Above transformations do not work when certain coefficients vanish or are equal to each other in~(\ref{eq:most_general_partial}), e.g $N=0$ and $P=L$. In these cases, solving~(\ref{eq:most_general_c_i}) is already much easier before any affine transformation of the arguments of $f$. In short, using a Gaussian ansatz allows us to solve any equation in the form of~(\ref{eq:most_general_partial}) without much trouble, as long as no nonlinear terms arise.

We will not discuss the transformations of the coefficients in~(\ref{eq:most_general_partial}) under the affine transformations of the arguments of $f$, e.g $C\rightarrow C' \rightarrow C''$. They follow from basic algebra.

\end{document}